\documentclass[aps,prb,twocolumn,a4paper,10pt,showpacs,citeautoscript,preprintnumbers,nobibnotes]{revtex4-1}
\usepackage[english]{babel}	

\usepackage{amsmath}		
\usepackage{amssymb}		
\usepackage{bbm}		

\usepackage{graphicx}		
\usepackage{graphics}		
\DeclareGraphicsExtensions{.png,.pdf}
\usepackage{color}		

\usepackage[colorlinks=true, a4paper=true, pdfstartview=FitV,linkcolor=blue, citecolor=blue, urlcolor=blue]{hyperref}
\newcommand{\sro}{Sr$_2$RuO$_{4}$~}
\newcommand{\SRO}{\sro}
\newcommand{\ie}{{\it i.e.~}}
\newcommand{\eg}{{\it e.g.~}}
\newcommand{\resp}{\emph{resp.~}}

\newcommand{\Equation}[2]{\begin{equation}\label{#1}#2\end{equation}}
\newcommand{\Align}[2]{\begin{align}\label{#1}#2\end{align}}
\newcommand{\Eqref}[1]{\eqref{#1}}
\newcommand{\Figref}[1]{Fig.~\ref{#1}}

\renewcommand\Re{\mathrm{Re}}
\renewcommand\Im{\mathrm{Im}}
\newcommand{\bs}{\boldsymbol}

\newcommand{\oo}{{(1)}}
\newcommand{\ot}{{(2)}}
\newcommand{\oa}{{(a)}}
\newcommand{\ob}{{(b)}}

\newcommand{\psia}{\psi^{(a)}}
\newcommand{\psiac}{\psi^{(a)*}}
\newcommand{\psio}{\psi^{(1)}}
\newcommand{\psioc}{\psi^{(1)*}}
\newcommand{\psit}{\psi^{(2)}}

\begin{document}
\title{\texorpdfstring{Vortex coalescence and type-1.5 superconductivity in \SRO}{Vortex coalescence and type-1.5 superconductivity in Sr2RuO4}}
\author{
 Julien Garaud${}^{1,2}$, Daniel F.~Agterberg${}^3$, Egor Babaev${}^{1,2}$}
\affiliation{
${}^1$Department of Physics, University of Massachusetts Amherst, MA 01003 USA \\
${}^2$Department of Theoretical Physics, The Royal Institute of Technology, Stockholm, SE-10691 Sweden \\
${}^3$Department of Physics, University of Wisconsin-Milwaukee, Milwaukee, WI 53211, USA
}

\begin{abstract}
Recently vortex coalescence was reported in superconducting  \SRO by several experimental
groups for fields applied along the $c$-axis.  We argue that \SRO is a type-1.5 superconductor
with long-range attractive, short-range repulsive intervortex interaction.
The type-1.5 behavior stems from an interplay of the two orbital degrees of freedom describing
this chiral superconductor  together with the  multiband nature of the superconductivity. These
multiple degrees of freedom give rise to multiple coherence lengths, some of which are larger and
some smaller than the magnetic field penetration length, resulting in nonmonotonic intervortex
forces.
\end{abstract}

\maketitle

The superconducting state of strontium ruthenate is of great interest because of experimental
evidence that it is a chiral spin-triplet superconductor. Support for this point of view came 
through a variety of measurements that include: the complete suppression of the superconducting
transition temperature ($T_c$) with non-magnetic impurities \cite{Mackenzie}; NMR Knight shift 
measurements that show no change in the spin susceptibility with temperature in the superconducting 
phase \cite{ishida,murakawa}; muon spin measurements ($\mu$SR) that suggest broken time-reversal 
symmetry in the superconducting state \cite{luke}; polar Kerr effect showing the broken time-reversal 
symmetry and the presence of chirality in the superconducting phase \cite{xia}; and phase sensitive 
measurements consistent with a chiral spin-triplet state \cite{liu,kidwingira}. While these measurements 
provide a strong case for a chiral spin-triplet superconducting phase, the case is not iron clad. In 
particular, sizable edge currents are expected to flow at sample boundaries and at domain walls between
domains of opposite chirality \cite{matsumoto}. A search for these currents has been carried out
and they have not been observed \cite{moler1,moler2,science}. Additionally, for fields in the basal plane,
two superconducting phases are predicted \cite{Agterberg:98} and the corresponding phase
transition between these phases has also not been observed. Finally, if spin-orbit coupling is
sufficiently large, then the spin susceptibility should show no change for fields in the basal plane,
but is expected to decrease with temperature for fields along the $c$-axis. NMR data shows no
change on the spin susceptibility for fields both in the basal plane \cite{ishida} and along the
$c$-axis \cite{murakawa}.  In spite of these puzzles in interpreting the superconducting state as
a chiral spin-triplet superconductor, the evidence in support of this state remains strong and current
research is focused on addressing these puzzles within this framework.

Another interesting aspect of the \sro is the multiband nature of its superconductivity. The Fermi
surface of \sro contains three sheets of cylindrical topology labeled $\alpha,\beta$, and $\gamma$
\cite{Bergemann}. These sheets stem from Ru $d_{xz},d_{yz}$ and Ru $d_{xy}$ orbitals. In
particular, the $\alpha$ and $\beta$ sheets originate from the $d_{xz}$ and $d_{yz}$ orbitals which
lead to quasi-one-dimensional bands, and the $\gamma$ sheet originates from the $d_{xy}$ orbitals
which leads to a quasi-two-dimensional band. It has been shown that for any non $s$-wave pairing
state, the superconducting order on the $\gamma$ sheet is weakly coupled to that on the
$\alpha,\beta$ sheets \cite{Agterberg.Rice.ea:97,Zhitomirsky.Rice:01,Annett.Litak.ea:02}. This
enables the physical picture of a {multi} gap superconductor, where a full gap exists on the active
band (either the $\gamma$ or the $\alpha,\beta$ bands) and the gap on the passive band is driven
by the active gap through the weak coupling between the two gaps. The gap on the passive band
may or may not contain nodes depending upon the details of the coupling between the two bands
\cite{Agterberg.Rice.ea:97,Zhitomirsky.Rice:01}. This scenario is supported by the evolution of the
specific heat with magnetic field where it is seen that a small magnetic field is sufficient to remove
the gap in passive band \cite{Deguchi.Mao.ea:04}. The identification of the active band has been
a matter of debate. Specific heat measurements suggest that it is the $\gamma$ band
\cite{Deguchi.Mao.ea:04}. However, it has been recently argued that if the $\alpha,\beta$ bands
are the active band, then it is possible that the edge currents are not large
\cite{Raghu.Kapitulnik.ea:10,Imai.Wakabayashi.ea:12}, which provides a reason for why they have
not been observed (note that small edge currents may make it difficult to explain the fields seen 
by $\mu$SR measurements \cite{Ashby.Kallin:09}). Furthermore, unlike the $\gamma$ band, 
superconductivity in the $\alpha,\beta$ bands gives rise to an intrinsic Hall effect that naturally 
accounts for the observed polar Kerr effect \cite{Taylor.Kallin:12}.

A striking feature which has been  repeatedly observed in this material and is the subject of this
paper, is vortex coalescence into clusters in various samples
\cite{dolocan1,moler1,*dolocan2,moler2,bending}. The experimental works
Refs.~\onlinecite{dolocan1,moler1,*dolocan2,moler2} interpreted this as originating from attractive
intervortex interactions of unknown nature. Indeed the  mechanisms for possible small intervortex
attractions in single-component superconductors cannot apply for this material. In single-component
systems vortices have either purely repulsive (in type-2 case) or purely attractive (in type-1 case) 
interaction at the level of Ginzburg-Landau theory. The situation can be more complex in microscopic
models of weak-coupling type-2 superconductors with the Ginzburg--Landau parameter $\kappa$
extremely close to the Bogomolnyi limit $1/\sqrt{2}$. There, intervortex forces in Ginzburg--Landau
theory are vanishingly small. The long-range intervortex forces are  then determined by non-universal
microscopic physics, which for certain materials can lead to tiny attractive forces
\cite{jacobs,*leung1,*leung2,*klein} which  at long range can win over repulsive
interaction.  However this mechanism can be ruled out in the case of \sro
since the measurements of critical fields suggest that an estimate for a putative Ginzburg--Landau
parameter is too far outside the regime where these effects takes place: $\kappa \approx 2.6$
(see \eg Ref.~\onlinecite{moler2}).
There exists a  different mechanism for intervortex attraction specific for multicomponent systems.
Multicomponent superconductors can have  several coherence lengths $\xi_a$. In this case, the
intervortex interaction can be long-range attractive and short-range repulsive because of the
interplay of the multiple fundamental length scales of the theory. That is, a nonmonotonic interaction
occurs if the London penetration length $\lambda$ falls between the coherence lengths
$\xi_1< {\lambda} < \xi_2$
\cite{bs11,*johan1,johan2,silaev1,moshchalkov,*moshchalkov2,silaev2,nonpairwise,dao,geurts}.
This regime was recently termed ``type-1.5" superconductivity\cite{moshchalkov,*moshchalkov2}
since it features coexisting and competing type-I and type-II behaviors. The nonmonotonic intervortex
forces originate there from the ``double-core" structure of vortices due to multiple coherence lengths.
The outer core (roughly speaking associated to the length scale $\xi_2$) extends outside the flux
carrying area. The overlap of the outer cores of vortices is responsible for the attractive intervortex
forces. Yet in the type-1.5 regime the vortices have short range   repulsion (due to current-current
and electromagnetic interaction) and are thermodynamically stable\cite{bs11}. This should lead to
vortex cluster formation in low magnetic field.
Here we examine the role of the multiple superconducting degrees of freedom that exist in \SRO.
After it was suggested in Ref.~\onlinecite{moshchalkov,*moshchalkov2} that type-1.5 regime is
realized in MgB$_2$~, a question was raised in the recent experimental work on \sro \cite{moler2}
whether or not the vortex coalescence in \sro originates via a type-1.5 scenario. Here we argue that for
realistic choices of phenomenological parameters, vortex clustering and type 1.5 behavior does
occur in multiband chiral Ginzburg--Landau theories for \SRO.

We consider a two-band chiral Ginzburg--Landau theory which incorporates the multiband nature
of the superconductivity. In accordance with the discussion above, the superconducting gap function
is taken to belong to the $E_u$ representation of the tetragonal point group. This representation
describes a chiral superconductor. The four complex components of the order parameter are
related to the spin-triplet gap functions by
${\bf d}_a({\bf k})=[\psia_1 f_{x,a}({\bf k})+\psia_2 f_{y,a}({\bf k})]\hat{z}$ where the $a$ labels
one of the two bands and the functions $f_{x,a}({\bf k})$ and $f_{y,a}({\bf k})$ share the same
symmetry properties as $k_x$ and $k_y$ under the symmetry operations of the point group
$D_{4h}$. In the units where $\hbar=1,c=1,m=1$, the two-band situation can be modeled as follows:
\Align{freeEnergy}{
 \mathcal{F}&= |\nabla\times\bs A|^2+															\nonumber \\
 &+\sum_{a=1,2}\Biggl\lbrace\delta_a\left( |D_x\psia_1|^2+ \gamma_a |D_y\psia_1|^2 \right. \nonumber \\
&\left.+ |D_y\psia_2|^2+ \gamma_a |D_x\psia_2|^2 \right)
	\nonumber \\
 &+2\delta_a\gamma_a\Re\left[ (D_x\psia_1)^*D_y\psia_2+ (D_y\psia_1)^*D_x\psia_2 \right]				\nonumber \\
 &+\beta_a\gamma_a\Re(\psia_1\ ^{*2}\psia_2\ ^2) 	 +\beta_a(2\gamma_a-1)|\psia_1|^2|\psia_2|^2		\nonumber \\
 &+\sum_{b=1,2}\alpha_a|\psia_b|^2+\frac{\beta_a}{2}|\psia_b|^4	\Biggl\rbrace						\nonumber \\
 &+2\nu\sum_{b=1,2}\Re\left[ \psioc_b\psit_b \right]					 \,,
}
where $\psia_b$ represent the superconducting components in the different bands, $a=1,2$
denotes the band index, $b=1,2$ denotes the two different components of each condensate.
$\Re[]$ stands for real part of the expression in brackets. Each component of a given condensate is a
complex function $\psi_a=|\psi_a|\exp\lbrace i\varphi_a\rbrace$. The gauge covariant derivative
is $\bs D=\nabla+ie\bs A$ and $\bs B = \nabla\times \bs A$. The gauge coupling constant $e$ is 
used to parametrize the penetration length of the magnetic field. This free energy is comprised of
two free energies, one for each band, that are coupled  through the vector potential ${\bf A}$
and by the parameter $\nu$. The free energies of each band are determined by weak coupling
theory in the clean limit (both are reasonable assumptions for \sro).  No assumptions are made
about the properties or geometry of the Fermi surface. The superconducting gap functions
$f_{y,a}({\bf k})$ are assumed to be given by the Fermi velocity components $v_x({\bf k})$ and
$v_y({\bf k})$. This approximation provides the correct momentum dependence of the gap for the
limiting cases of a purely cylindrical Fermi surface and a perfectly square Fermi surface
\cite{Agterberg:98}. The direct coupling of the bands through the sole parameter $\nu$ is also
justified within a weak coupling theory. The phenomenological theory has nine parameters that
require specification. Experiments place constraints on these parameters. In particular, one band
is required to be passive and the other active. We assume that the $\gamma$-band is the active
band and that the $\alpha/\beta$ bands give rise to a passive gap.
This assumption is not critical, the results will be similar if we assume the other possibility
\cite{Raghu.Kapitulnik.ea:10,Imai.Wakabayashi.ea:12,Taylor.Kallin:12}. This assumption implies
$\alpha_1<0$ and $\alpha_2>0$. Since the realization of type-1.5 state will be very similar
for   $\alpha_1>0$ and $\alpha_2<0$, without loss of generality we
report numerical investigation of only the first scenario.

 Additionally, for magnetic fields applied along the $c$-axis, the
free energy should reproduce the experimentally determined ratio of $H_{c2}$ and $H_{c1}$.
Finally, weak coupling calculations show that $\gamma_1>1/3$ and $\gamma_2<1/3$ and
consistency with the small observed anisotropy of the in-plane upper critical field requires that
both $\gamma_1$ and $\gamma_2$ are close to $1/3$ \cite{daniel2001}.

Equations of motion of the gauge field defines the supercurrent
\Equation{2B-Supercurrent1}{
 \bs J= \sum_{a, b} \bs J^\oa_\ob \,,
}
where the contribution of each component of a given condensate is
\Align{2B-Supercurrent2}{
 J^\oa_{\oo,x} & = \frac{e\delta_a}{2}\Im\left[\psiac_1(D_x\psia_1+\gamma_a D_y\psia_2) \right] 			\nonumber \\
 J^\oa_{\oo,y} & = \frac{e\delta_a\gamma_a}{2}\Im\left[\psiac_1(D_y\psia_1+ D_x\psia_2) \right] 			\nonumber \\
 J^\oa_{\ot,x} & = \frac{e\delta_a\gamma_a}{2}\Im\left[\psiac_2(D_x\psia_2+ D_y\psia_1) \right] 			\nonumber \\
 J^\oa_{\ot,y} & = \frac{e\delta_a}{2}\Im\left[\psiac_2(D_y\psia_2+\gamma_a D_x\psia_1) \right] 			\,,
}
where $\Im$ stands for imaginary part.

\begin{figure}[!htb]
 \hbox to \linewidth{ \hss
 \includegraphics[width=\linewidth]{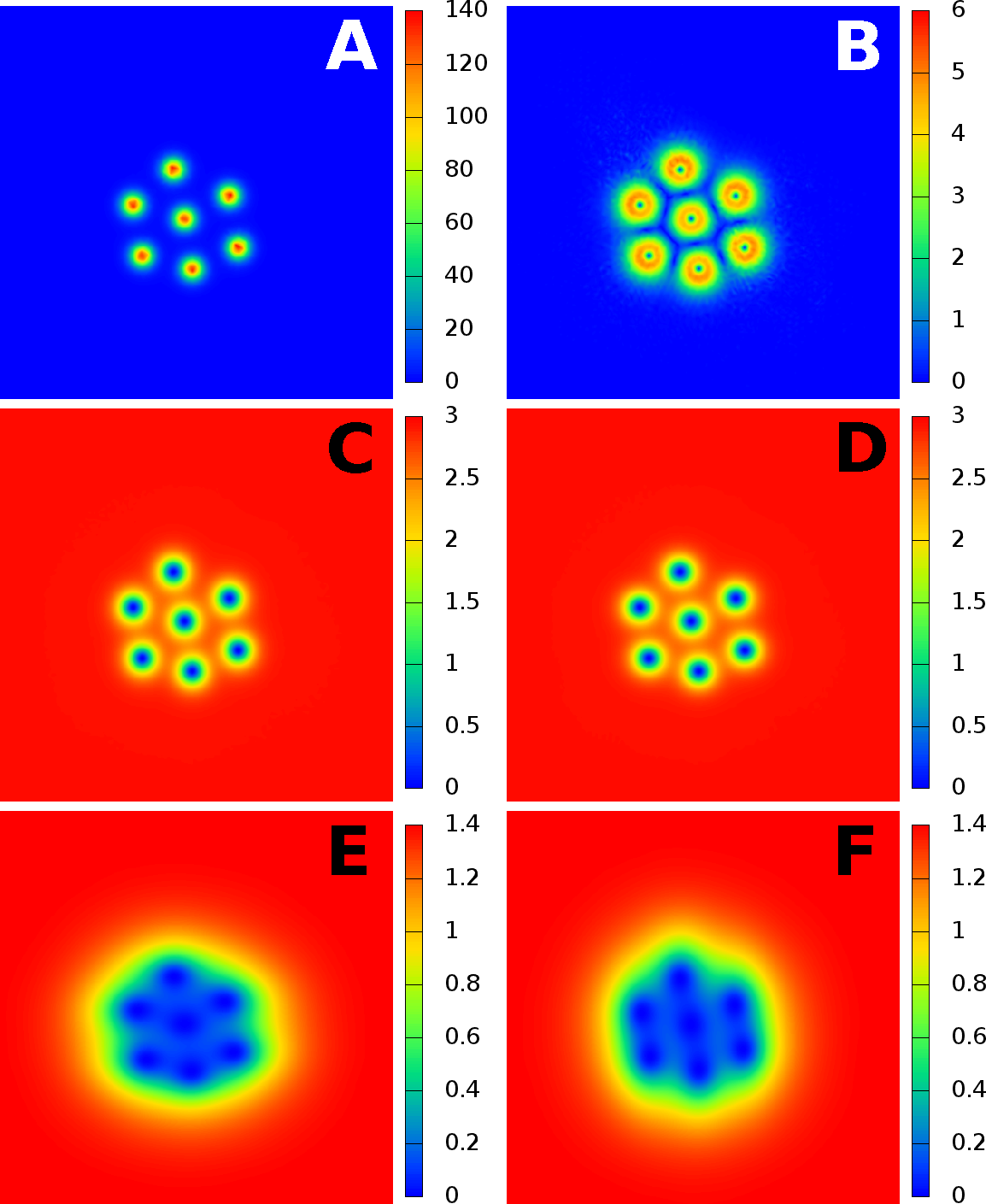}
\hss}
\caption{(Color online) --
A vortex cluster obtained numerically by energy minimization from a dilute system of seven
vortices. Displayed quantities are the magnetic field $B_z$ on $(\bf A)$ and the total supercurrent
\Eqref{2B-Supercurrent1} on $(\bf B)$. The densities of the first (\resp second) component
of the condensate $\psio$, namely $|\psio_1|^2$ (\resp $|\psio_2|^2$), are shown on $(\bf C)$
(\resp $(\bf D)$). The panels $(\bf E)$ (\resp $(\bf F)$) show the densities of the first
(\resp second) component of the condensate $\psit$, that is $|\psit_1|^2$ (\resp $|\psit_2|^2$).
The figure demonstrates existence of several length scales associated with the density
variations. The panels $(\bf E)$ (\resp $(\bf F)$) show extended cores, the overlap of these
cores leads to attractive intervortex forces.
}
\label{Fig:Cluster}
\end{figure}

Coalescence of vortex matter into clusters is investigated numerically by minimizing the
free energy \Eqref{freeEnergy} within a finite element framework provided by the
FreeFem++ library \cite{FreefemNote}. We investigated vortex matter in the model 
\Eqref{freeEnergy} for parameters which give characteristic length scales which are close 
to experimental estimates for \sro. We report the case  $(\alpha_1,\beta_1)=(-10,10)$, 
$\gamma_1=0.35$ and $\delta_1=1$. Parameters associated with the second condensate 
are $(\alpha_2,\beta_2)=(0.3,1)$, $\gamma_2=0.25$ and $\delta_2=1$. The interband coupling 
is $\nu=0.45$ and the electric charge $e=1.8$. We also investigated a range of similar parameters 
to ensure that they give a similar picture (i.e. the parameter set does not correspond to any 
fine-tuned situation). For a general study how type-1.5 behavior is affected by interband coupling 
$\nu$ see Refs.~\onlinecite{johan1,johan2,silaev1,silaev2}.
Note that,  the models like \Eqref{freeEnergy} can  have a Skyrmionic phase, where the vortices are 
unstable against a decay to Skyrmions \cite{garaudskyr}. We obtained stable vortex clusters in various 
type-1.5 regimes which are stable against a decay into Skyrmions. \Figref{Fig:Cluster} shows a numerical 
solution for one such vortex cluster. We also find that Skyrmions can still exist in those regimes as
metastable topological excitations, which are  more energetically expensive than vortices.

Note that the experiments \cite{moler1,dolocan1,*dolocan2,moler2} were performed at
$T\ll T_c$. A microscopic approach is therefore required
to get all aspects of the physics quantitatively correct since the Ginzburg--Landau model is
quantitatively correct only for $T$ near to $T_c$. However, under certain circumstances, a
phenomenological Ginzburg--Landau model can qualitatively and quantitatively describe the
long-range intervortex forces in the type-1.5 regime even at relatively low temperatures. This
was shown in a two-band system with not too strong interband coupling and one passive band
\cite{silaev2}. The physics that a  GL-based approach fails to describe in the low temperature
regime, is primarily associated with shorter-length scales  such as the counterpart of Kramer-Pesch
effect. Since in our study we are only interested in long-range intervortex forces (\ie we do not
consider the high-field regimes, where intervortex distance is short), the phenomenological GL
approach still provides a qualitatively correct picture of that physics even at relatively low
temperatures.

\begin{figure}[!htb]
 \hbox to \linewidth{ \hss
 \includegraphics[width=0.6\linewidth]{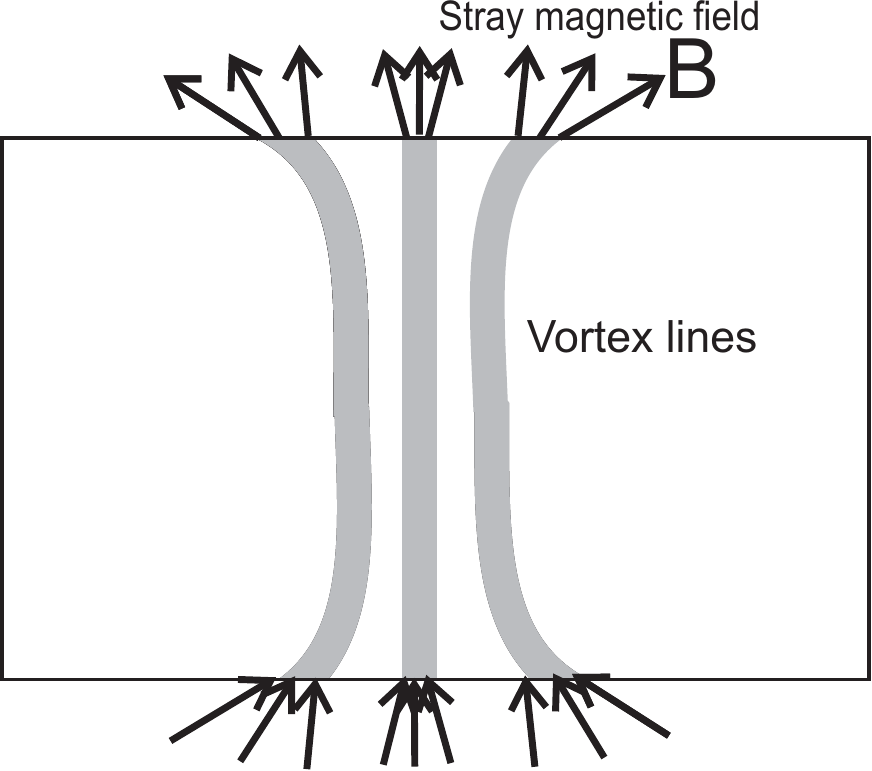}
 \hss}
\caption{
Even if vortices have attractive forces in the bulk, their segments near the surface can have
long-range contributions to repulsive interaction due to the effects of demagnetization fields
outside the sample. This is schematically shown on this figure. Thus a surface probe of a bulk
sample can overestimate the position of the minimum in the intervortex potential. Also in case
of a thin film, the stray fields can make intervortex distance larger.
}
\label{broadening}
\end{figure}

The experiments in Refs.~\onlinecite{dolocan1,*dolocan2}, \onlinecite{moler2} and
\onlinecite{bending} obtain rather similar intervortex distances within the clusters. The typical
intervortex distance is larger than their estimates of the London penetration length. Intervortex
attraction is indeed achievable in type-1.5 regime at such length scales. The minimum of the
intervortex interaction potential is determined not only by the length scales $\lambda,\xi_1,\xi_2$
but also by non-linear effects. Thus in a type-1.5 regime intervortex distance in a cluster could
be substantially larger than the London penetration length scale. For example, the intervortex
potential shown on Figure 2 in Ref.~\onlinecite{johan2} corresponds to such a situation.
It should be noted that all experiments observing vortex clusters
(Refs.~\onlinecite{dolocan1,moler1,*dolocan2,moler2}), are scanning SQUID or Hall experiments
which probe magnetic field at a small distances over the sample's surface. It should be kept in mind
that such a surface probe could in general overestimate the position of the minimum of the
interaction potential, for vortices with nonmonotonic interactions. Indeed, near the surface, the
long-range intervortex forces can be altered by the electromagnetic repulsion caused by the stray
fields outside the sample (schematically shown in \Figref{broadening}). Similarly, stray field physics
affects the structure of normal domains in ordinary type-1 superconductors \cite{Gennes}.

The simplest origin of inhomogeneous vortex distributions are the pinning effects.
And in fact in some of the samples, existence of preferential pinning areas were
identified in the work of Curran {\it et. al.} \cite{bending,bending2}, see also remark 
 \footnote{Note that in chiral superconductors vortex pinning can arise not only
from sample inhomogeneities but also due to vortex interaction with pinned domain
walls between  different chiral domains \cite{machida,*vakaryuk,*goldbart,garaudskyr} . }.
However, in the experimental papers \cite{dolocan1,*dolocan2} it was argued that the 
observed vortex clustering is not related to pinning. The layered structure of \sro leads to 
vortex stripes formation when in-plane field is applied. Therefore it is possible to move vortices 
by applying the in-plane component of the external magnetic field. Thus it is possible to assess 
this way, if the vortex clusters are artifacts of some local pinning landscape. In the
Ref.~\onlinecite{dolocan1,*dolocan2} it was argued that the mobility of the  vortex clusters
in their samples is inconsistent with clustering due to a pinning scenario.

The distribution of the intervortex distance in several  samples with a relatively small number of 
vortices was analyzed in  \cite{bending}. They found that it did not exhibit a clear peak at certain 
preferred intervortex distance, which was interpreted as being inconsistent with the type-1.5 scenario. 
We argue however that it does not necessarily contradict the
type-1.5 scenario, for the following reasons. The shape of the vortex clusters in a certain subset
of type-1.5 regimes is  rather substantially affected by non-pairwise contributions to intervortex
forces \cite{nonpairwise} (and also by dynamic and entropic aspects associate with it). This can
affect the distribution of intervortex distances. Since the multiband model of \sro has four
components, the non-pairwise contributions to intervortex forces could be relatively significant
and produce variation in the shapes of vortex clusters. Due to these effects, it can be  more difficult
to get a distinguishable double-peak structure from vortex distributions in small samples like those
studied in Ref.~\onlinecite{bending}. Large samples with a larger number of vortices can yield a
more conclusive answer.
The absence of collapse to a single vortex cluster noticed in Refs.~\onlinecite{dolocan1,*dolocan2}
and \onlinecite{moler2}, also naturally arises in type-1.5 regime. Again it can originate in dynamic
and entropic reasons and be enhanced by non-pairwise contributions to intervortex forces
\cite{nonpairwise}. It can also be a result of demagnetization (stray fields) effects.

In conclusion, we have shown that vortex coalescence occurs in a realistic phenomenological
model describing the multicomponent superconductivity in \SRO.  This provides an explanation
for the vortex coalescence that was reported by several experimental groups. These experiments
can be interpreted as demonstrating that \sro is a type-1.5 superconductor: \ie it possess several
coherence length: some longer than the London penetration length and some shorter. In order to
firmly establish this interpretation, further experimental studies on the distribution of vortices are
required.

We thank Peter Curran, Clifford Hicks, Simon Bending, Kam Moler and Victor Moshchalkov for
useful correspondence. This work was supported by Knut and Alice Wallenberg Foundation through
the Royal Swedish Academy of Sciences, Swedish Research Council and by the NSF CAREER Award
No. DMR-0955902. DFA is supported by NSF grant DMR-0906655. The computations were performed
on resources provided by the Swedish National Infrastructure for Computing (SNIC) at National
Supercomputer Center at Link\"oping, Sweden.

%

\end{document}